\documentclass[12pt]{article}
\usepackage{epsfig}
\usepackage{graphicx}
\usepackage{amssymb}
\usepackage{amsmath}
\usepackage{amsfonts}
\newcommand{\bc}{\begin{center}}
\newcommand{\ec}{\end{center}}
\newcommand{\R}{\mathbb{R}}
\newcommand{\C}{\mathbb{C}}

\newcommand{\Z}{\mathbb{Z}}

\newcommand{\fg}{\mathfrak{g}}
\newcommand{\fC}{\mathfrak{C}}

\newcommand{\fA}{\mathfrak{A}}
\newcommand{\fB}{\mathfrak{B}}
\newcommand{\fS}{\mathfrak{S}}

\newcommand{\fI}{\mathfrak{I}}

\newcommand{\fM}{\mathfrak{M}}

\newcommand{\fP}{\mathfrak{P}}

\newcommand{\fa}{\mathfrak{a}}
\newcommand{\fb}{\mathfrak{b}}
\newcommand{\fp}{\mathfrak{p}}

\newcommand{\fu}{\mathfrak{u}}

\newcommand{\fH}{\mathfrak{H}}

\newcommand{\be}{\begin{equation}}
\newcommand{\ee}{\end{equation}}
\newcommand{\bea}{\begin{eqnarray}}
\newcommand{\eea}{\end{eqnarray}}

\newcommand{\nn}{\nonumber}
\newcommand{\kt}{\rangle}
\newcommand{\br}{\langle}

\newcommand{\ed}{\end{document}}

\newcommand{\cH}{\mathcal{H}}
\newcommand{\cN}{\mathcal{N}}
\newcommand{\cK}{\mathcal{K}}

\newcommand{\cG}{\mathcal{G}}
\newcommand{\cR}{\mathcal{R}}

\newcommand{\cM}{{\cal M}}
\newcommand{\cP}{{\cal P}}

\newcommand{\cC}{\mathcal{C}}
\newcommand{\cA}{\mathcal{A}}

\newcommand{\cU}{\mathcal{U}}

\newcommand{\bpi}{{\mbox{\large $\pi$}}}

\begin{document}

\title{Geometric Phase for Non-Hermitian Hamiltonians\\ and Its
Holonomy Interpretation\footnote{Published as J.~Math.~Phys.\
\textbf{49}, 082105 (2008).}}

\author{Hossein Mehri-Dehnavi\thanks{E-mail address:
mehrideh@iasbs.ac.ir} ~and Ali~Mostafazadeh\thanks{E-mail address:
amostafazadeh@ku.edu.tr} \\
\\
$^*$~Department of Physics, Institute for Advanced Studies in
Basic \\ Sciences, Zanjan 45195-1159, Iran \\
$ ^\dagger$~Department of Mathematics, Ko\c{c} University,
Rumelifeneri Yolu,\\ 34450 Sariyer, Istanbul, Turkey }
\date{ }
\maketitle

\begin{abstract}
For an arbitrary possibly non-Hermitian matrix Hamiltonian $H$, that
might involve exceptional points, we construct an appropriate
parameter space $\fM$ and a lines bundle $L^n$ over $\fM$ such that
the adiabatic geometric phases associated with the eigenstates of
the initial Hamiltonian coincide with the holonomies of $L^n$. We
examine the case of $2\times 2$ matrix Hamiltonians in detail and
show that, contrary to claims made in some recent publications,
geometric phases arising from encircling exceptional points are
generally geometrical and not topological in nature.\vspace{5mm}

\vspace{2mm}

\end{abstract}
\vspace{5mm}

\noindent



\section{Introduction}

A state of a quantum system is said to be cyclic, if its dynamical
evolution yields the same state after a time $T$. For a pure state,
this means that a corresponding state vector returns to the same ray
in the Hilbert space, but acquires a phase factor. The discovery
that this phase factor consists of a dynamical and a geometrical
part has important implications \cite{gp-book}. The first general
treatment of geometric phases that underlined their significance and
made them a focus of intensive research in the past three decades is
due to Berry \cite{berry} who considered Hermitian Hamiltonians
undergoing adiabatic changes. Berry's adiabatic geometric phase
admits a variety of generalizations. Among these are non-Abelian
\cite{WZ}, non-adiabatic \cite{aharonov}, non-cyclic
\cite{bhandari,jpa-1999}, classical \cite{classical-gp}, and
relativistic \cite{jpa-1998} geometric phases. There have also been
attempts to extend geometric phases to systems described by
non-Hermitian Hamiltonians \cite{gari}.\footnote{An extensive list
of references on geometric phases and their generalizations that
appeared before the year 2000 is given in \cite{nova}.}

Geometric phases for systems having a Hermitian Hamiltonian have an
interesting mathematical interpretation in terms of the holonomies
of certain fiber bundles \cite{gp-book,aharonov,simon}. The main
purpose of the present article is to offer a comprehensive holonomy
interpretation for the adiabatic geometric phases associated with
non-Hermitian matrix Hamiltonians, particularly those involving
exceptional points. It is known that in the presence of such points,
an adiabatic periodic change of the Hamiltonian does not necessarily
imply a cyclic evolution of its eigenstates. Even for the simple
$2\times 2$ Hamiltonians that have widely been investigated in
recent years \cite{his-prl,mail1}, one often needs to evolve the
initial state vector for two complete periods of the Hamiltonian to
achieve a cyclic evolution. This behavior makes a direct
generalization of the results of \cite{simon} for general
non-Hermitian Hamiltonians intractable. In this article we offer a
complete solution for the problem of the construction of the
relevant parameter spaces and the fiber bundles that allow for a
holonomy interpretation of the adiabatic geometric phases for
general non-Hermitian matrix Hamiltonians. We also address the
controversy related with the topological nature of these phases.

A significant difference between Hermitian and non-Hermitian
Hamiltonians is the nature of their degeneracies. Hermitian
Hamiltonians are diagonalizable, i.e., they admit a complete set of
eigenvectors. This is true even at a degeneracy point where two or
more eigenvalues coalesce. In contrast, for a non-Hermitian
Hamiltonian, in addition to the usual degenerate points where the
Hamiltonian is diagonalizable, there may exist points of degeneracy
where one can no longer form a complete set of eigenvectors. We will
respectively refer to these two types of degeneracy points as
\textit{diabolic degeneracy points} and \textit{exceptional
degeneracy points}. Following the terminology employed in
\cite{heiss-jpa-1990} we will abbreviate the latter as
\emph{exceptional points} (EP).\footnote{Kato \cite{kato}, from whom
the term ``exceptional point'' seems to have been adopted,
identifies an exceptional point with a one at which the number of
distinct eigenvalues of the operator in question changes. For
example, according to Kato, $x=1$ is an exceptional point of the
$2\times 2$ Hamiltonian $H[x]={\rm diag}(x,1)$. In our
classification this point is a diabolic degeneracy point, where none
of the interesting phenomena reported in the related literature
\cite{his-prl,mail1,his-chez,his-jpa,gunther} occurs. In
\cite{berry-chez}, EPs are called non-Hermitian degeneracies.}
Non-Hermitian Hamiltonians with EPs appear in the study of
resonances \cite{mondragon,heiss-pre-20000}, quantum chaos
\cite{heiss-jpa-1990,heiss-pr-1994}, quantum cosmology
\cite{jmp2002d,ap-2004}, magneto-hydrodynamics
\cite{gunther-cjp-2004}, and various classical and quantum systems
\cite{mail1,his-chez,his-jpa,gunther,berry-chez,stehmann-jpa-2004,rotter-pre-2003,heiss-pra-1998}.
Geometric phases associated with EPs have been the subject of
various theoretical \cite{mail1,gunther,nesterov,keck} and
experimental \cite{his-prl,his-pre} studies.

A controversial claim about the geometric phases associated with EPs
is that they have a topological character \cite{mail1}. This means
that these phases are invariant under smooth perturbations of the
path traversed by the parameters of the system. The main
characteristic feature of topological phases is that the phase angle
is an integer multiple of $\pi$, i.e, the phase factor is merely a
sign. This is the case for the adiabatic geometric phases for a real
symmetric Hamiltonian \cite{gp-book,berry,higgins,mead-truhlar}. We
will argue that the claim pertaining the topological nature of
EP-related geometric phases \cite{mail1} rests on the specific
choice of the path of parameters made in the previous studies of the
subject, and that it does not generally hold.

Some authors restrict their study of EP-related geometric phases to
systems having a complex symmetric Hamiltonian \cite{gunther,keck}.
Often they claim that this restriction does not diminish the
generality of their results, because every square matrix is related
to a symmetric matrix by a similarity transformation. One must
however note that the similarity transformations that map an
arbitrary time-dependent Hamiltonian into a symmetric one are
generically time-dependent. These time-dependent similarity
transformations induce an additional term in the Schr\"odinger
equation that in general changes the value of the geometric phases
and other physical quantities. Therefore, contrary to the
above-mentioned claim, a complete treatment of EP-related geometric
phases calls for the consideration of general possibly nonsymmetric
Hamiltonians.\footnote{The same is true about Hermitian
Hamiltonians. Although every Hermitian Hamiltonian can be mapped via
a similarity transformation to a real symmetric (in fact real
diagonal) matrix, to investigate geometric phases for Hermitian
Hamiltonians one cannot confine one's attention to real symmetric or
diagonal Hamiltonians.}

In \cite{sun} the author discusses a straightforward extension of
the approach of \cite{simon} to offer a holonomy interpretation of
the adiabatic geometric phases for the non-Hermitian matrix
Hamiltonians considered in \cite{gari}. The treatment of \cite{gari}
and \cite{sun} is based on the assumption that the eigenvalues of
the Hamiltonian are single-valued functions of its parameters. This
holds typically whenever the path of parameters of the Hamiltonian
does not enclose an EP. The fact that this is not generally the case
is at the root of the difficulties associated with devising a
consistent holonomy interpretation of the adiabatic geometric phases
for a general non-Hermitian Hamiltonian.

The organization of the article is as follows. In Section~2, we
outline the construction of the parameter spaces for quantum systems
defined by non-Hermitian matrix Hamiltonians. In Section~3, we
construct the desired line bundle over the relevant parameter space.
In Section~4, we calculate the adiabatic geometric phases for a
general $2 \times 2$ complex matrix Hamiltonian and examine various
specific examples. Here we also discuss the topological versus
geometric nature of the EP-related geometric phases. Finally, in
Section~5 we summarize our results and present our concluding
remarks. The article includes two appendices that are respectively
devoted to an algebraic topological treatment of the parameter
spaces constructed in the text and a derivation of an explicit
expression for the adiabatic geometric phase for the cases that the
curve in the parameter space does not lie in a single open patch but
finitely many patches.

\section{Construction of the Parameter Space}

Consider an arbitrary $N\times N$ matrix Hamiltonian $H$. We can
parameterize $H$ in terms of its $N^2$ complex entries. These
correspond to $2N^2$ real numbers $(X_1,X_2,\cdots,X_{2N^2})=:X$
that we can identify with the elements of $\R^{2N^2}$. We will use
the symbol $H[X]$ for the Hamiltonian to mark its $X$-dependence.
The condition that $H[X]$ has $N$-distinct eigenvalues restricts $X$
to a connected open subset $M_{\rm max}$ of $\R^{2N^2}$ that
inherits the structure of a smooth $2N^2$-dimensional real manifold
from $\R^{2N^2}$.

Let $M$ be a $d$-dimensional connected submanifold of $M_{\rm max}$
whose points $R$ are locally expressed by the coordinates
$(R_1,R_2,\cdots,R_d)$. We shall consider the cases that the
Hamiltonian is parameterized by the elements of $M$. We denote this
using $H[R]$ in place of $H[X]$. Clearly, the largest possible
choice for $M$ is $M_{\rm max}$ in which case $d=2N^2$. If we
consider a general Hermitian $N\times N$ matrix Hamiltonian, $d=N^2$
and $M$ is a submanifold of $\R^{N^2}$. For $N=2$, we have
$M=\R\times(\R^3-\{0\})=\R\times\R^+\times S^2$ where
$S^2:=\{(x,y,z)\in\R^3|x^2+y^2+z^2=1\}$.

Let $\fI$ be a (nonempty) indexing set,
$\{U_\alpha\}_{\alpha\in\fI}$ be an open cover of $M$ such that
$U_\alpha$ is diffeomorphic to $\R^d$ and there are single-valued
smooth functions
    \be
    E^\alpha_1:U_\alpha\to\C,~~~~E^\alpha_2:U_\alpha\to\C,~~~\cdots,~~~
    E^\alpha_N:U_\alpha\to\C,
    \label{eigenvalues}
    \ee
such that for all $R\in U_\alpha$ and all
$j\in\{1,2,\cdots,N\}=:I_N$, $E^\alpha_j[R]$ is an eigenvalue of
$H[R]$. Note that such an open cover of $M$ always exists
\cite{kato} and because $H[R]$ has a nondegenerate spectrum, it has
$N$ distinct eigenvalues; for all $m,n\in I_N$,
    \be
    E^\alpha_m[R]\neq E^\alpha_n[R]~~~~~~
    \mbox{if and only if}~~~~~~m\neq n.
    \label{nondegenerate}
    \ee
This implies the existence of a complete biorthonormal system
$\{|\psi_{j}^{\alpha}[R]\kt,|\phi_{j}^{\alpha}[R]\kt\}$ of the
Hilbert space $\cH$ that satisfies
    \be
    H[R]|\psi_{j}^{\alpha}[R]\kt
    =E_{j}^{\alpha}[R]|\psi_{j}^{\alpha}[R]\kt,~~~~
    H^{\dag}[R]|\phi_{j}^{\alpha}[R]\kt
    ={E^{*}}^{\alpha}_{j}[R]|\phi_{j}^{\alpha}[R]\kt,
    ~~~~\br\phi_{j}^{\alpha}[R]|\psi_{k}^{\alpha}[R]\kt
    =\delta_{j,k}.
    \label{biortho}
    \ee
Here $\alpha\in\fI$ and $j,k\in I_{N}$ are arbitrary, $\dag$ and $*$
respectively denote the adjoint and complex conjugate, and
$\delta_{m,n}$ stands for the Kronecker delta symbol. It is
important to note that $|\psi_{j}^{\alpha}[R]\kt$ and
$|\phi_{j}^{\alpha}[R]\kt$ define single-valued (smooth) functions
mapping $U_\alpha$ to $\cH$, namely
$|\psi_{j}^{\alpha}[\cdot]\kt:U_\alpha\to\cH$ and
$|\phi_{j}^{\alpha}[\cdot]\kt:U_\alpha\to\cH$.

Conditions (\ref{biortho}) do not determine the biorthonormal system
$\{|\psi_{j}^{\alpha}[R]\kt,|\phi_{j}^{\alpha}[R]\kt\}$ uniquely,
but they imply that any other biorthonormal system,
$\{|\psi_{j}^{'\alpha}[R]\kt,|\phi_{j}^{'\alpha}[R]\kt\}$,
fulfilling these conditions have the form
    \bea
    |{\psi_{j}'}^{\alpha}[R]\kt: =
    k_{j}^{\alpha}[R]|\psi_{j}^{\alpha}[R]\kt,~~~~
    |{\phi_{j}'}^{\alpha}[R]\kt:=
    \frac{1}{k_{j}^{\alpha}[R]^{*}}|\phi_{j}^{\alpha}[R]\kt.
    \label{gauge}
    \eea
Here $j\in I_N$ is arbitrary and
$k_{j}^{\alpha}:U_\alpha\to\C-\{0\}$ are smooth (single-valued)
functions.

By definition, the (pure) states of a quantum system are rays
(one-dimensional subspaces) of the Hilbert space $\cH$. Every state
is uniquely determined by any of its nonzero elements (state
vectors). We use $\lambda_\psi$ to denote the state determined by an
state vector $|\psi\kt\in\cH-\{0\}$, i.e.,
    \be
    \lambda_\psi:=\{c\,|\psi\kt~|~c\in\C~\}.
    \ee
States $\lambda_\psi$ are in one-to-one correspondence with the
projection operators
    \be
    \Lambda_{\psi}:=\frac{|\psi\kt
    \br\psi|}{\br\psi|\psi\kt}.
    \label{state}
    \ee
In particular, we can characterize the eigenstates
$\lambda_j^\alpha[R]$ of $H[R]$ that contain the eigenvectors
$|\psi_{j}^{\alpha}[R]\kt$ with the projection operators
    \be
    \Lambda_j^\alpha[R]:=\Lambda_{\psi_{j}^{\alpha}[R]}=\frac{|\psi_j^\alpha[R]\kt
    \br\psi_j^\alpha[R]|}{
    \br\psi_j^\alpha[R]|\psi_j^\alpha[R]\kt}.
    \label{eigenstate}
    \ee
Note that as expected the projection operators $\Lambda_j^\alpha[R]$
are invariant under the ``gauge transformations'':
$|{\psi_{j}}^{\alpha}[R]\kt\to|\psi_{j}^{'\alpha}[R]\kt$,
$|\phi_{j}^{\alpha}[R]\kt\to|\phi_{j}^{'\alpha}[R]\kt$,
\cite{gp-book}. This shows that the eigenstates
$\lambda_j^\alpha[R]$ define single-valued functions
$\Lambda_j^\alpha[\cdot]:U_\alpha\to {\cal L}(\cH)$, where ${\cal
L}(\cH)$ is the set of linear operators acting in $\cH$.

Similarly to the case of Hermitian Hamiltonians \cite{berry}, every
(parameterized) curve $C:[0,T]\to M$ defines a time-dependent
Hamiltonian according to $H(t):=H[R(t)]$, where $R(t)$ stands for
the coordinates of $C(t)$ in $U_\alpha$ and we have identified the
parameter $t\in [0,T]$ of the curve $C$ with time.\footnote{For
brevity we shall use the same label for the curve $C$ and its range
$\{C(t)|t\in[0,T]\}$.} We will consider the dynamics generated by
the Hamiltonian $H(t)$ via the Schr\"{o}dinger equation
    \bea
    H(t)|\Psi(t)\kt=i\frac{d}{dt}|\Psi(t)\kt,
    \label{scheq}
    \eea
where $|\Psi(t)\kt$ is an evolving state vector.

We say that $H(t)$ generates an \emph{adiabatic time-evolution}, if
the eigenstates $\lambda_j^\alpha[R(0)]$ of $H(0)$ evolve into the
eigenstates $\lambda_j^\alpha[R(t)]$ of $H(t)$ (with the same
spectral label $j$ and for all $t\in[0,T]$ and $\alpha\in\fI$),
i.e., given $\lambda_{\Psi(0)}=\lambda_j^\alpha[R(0)]$ we have
$\lambda_{\Psi(t)}=\lambda_j^\alpha[R(t)]$,
\cite{mondragon,sun,pla-1999}. It is not difficult to show that this
requirement puts too severe a restriction on the Hamiltonian
$H(t)$.\footnote{For Hermitian Hamiltonians it implies that the
eigenstates $\lambda_j^\alpha[R(0)]$ must actually be stationary
\cite{gp-book}.} Therefore, in practice, one requires that this
condition be satisfied approximately.

If $C$ is a simply closed curve, $H(t)$ is a periodic Hamiltonian
with period $T$. In addition, if $H(t)$ generates an adiabatic
time-evolution and $C$ lies entirely in $U_\alpha$ for some
$\alpha\in\fI$, then the eigenstates $\lambda_j^\alpha[R(0)]$ of
$H(0)$ are cyclic states and we may explore the corresponding
adiabatic geometric phases. However, if $C$ cannot be contained in a
single open patch of $M$, it may happen that for some $j\in I_N$,
$\alpha,\beta\in\fI$ and $C(t_\star)\in U_\alpha\cap U_\beta$,
$\lambda_j^\alpha[R(t_\star)]\neq \lambda_j^\beta[R(t_\star)]$.
Because of the non-degeneracy of the spectrum of the Hamiltonian,
this is equivalent to $E_j^\alpha[R(t_\star)]\neq
E_j^\beta[R(t_\star)]$. But since $E_j^\alpha[R(t_\star)]$ is an
eigenvalue of $H[R(t_\star)]$ regardless of whether we identify
$R(t_\star)$ with coordinates of $C(t_\star)$ associated with the
patch $U_\alpha$ or $U_\beta$, there should be some $k\in I_N-\{j\}$
such that $E_j^\alpha[R(t_\star)]=E_k^\beta[R(t_\star)]$. In other
words, it may happen that in certain points along the curve the
eigenvalues (and consequently the eigenstates) with different
spectral labels get swapped. If this happens
$\lambda_j^\alpha[R(0)]$ may no longer be a cyclic state. The
possibility of such a noncyclic adiabatic evolution has been ignored
in the earlier investigations of the problem
\cite{gari,mondragon,sun,pla-1999}. It was brought to light when
curves encircling EPs were considered.

The above-mentioned complication arise simply because the
eigenvalues of the parameter-dependent Hamiltonian are not
single-valued functions on the parameter space $M$. This calls for
the construction of appropriate parameter spaces for the problem.
This is actually necessary for devising a holonomy interpretation
for the adiabatic geometric phase of general non-Hermitian matrix
Hamiltonians. We shall construct the desired parameter spaces as
certain fiber bundles over $M$.

For each $\gamma\in\fI$, we define $s^\gamma:U_\alpha\to\C^N$
according to
    \be
    s^\gamma(R):=\left(\begin{array}{c}
    E^\gamma_1[R]\\ E^\gamma_2[R]\\
    \vdots\\
    E^\gamma_N[R]\end{array}\right),
    \label{phi}
    \ee
and let $\alpha,\beta\in\fI$ be such that $U_\alpha\cap
U_\beta\neq\varnothing$. For all $R\in U_\alpha\cap U_\beta$ and
$k\in I_N$, $E^\alpha_k[R]$ is an eigenvalue of $H[R]$. This implies
that there is a unique $j\in I_N$ such that
$E^\alpha_k[R]=E^\beta_j[R]$. We can use this observation to infer
the existence of a permutation $\sigma_R^{\beta,\alpha}\in \fS_N$
satisfying $j=\sigma_R^{\beta,\alpha}(k)$, where $\fS_N$ is the
permutation group of $I_N$.

$\sigma_R^{\beta,\alpha}$ relates $s^\alpha$ and $s^\beta$ according
to
    \be
    s^\alpha(R)\to s^\beta(R)=g(\sigma_R^{\beta,\alpha})
    s^\alpha(R),
    \label{phi-phi}
    \ee
where for every $\sigma\in\fS_N$, $g(\sigma)\in GL(N,\C)$ is the
$N\times N$ matrix defined by
    \[ g(\sigma)\left(\begin{array}{c}
    z_1 \\ z_2 \\
    \vdots\\
    z_N \end{array}\right)=\left(\begin{array}{c}
    z_{\sigma(1)} \\ z_{\sigma(2)}\\
    \vdots\\
    z_{\sigma(N)} \end{array}\right),~~~~~~~\mbox{for all}~~~
    \left(\begin{array}{c}
    z_1 \\ z_2 \\
    \vdots\\
    z_N \end{array}\right)\in\C^N,\]
i.e., $g:\fS_n\to GL(n,\C)$ gives the standard matrix representation
of $\fS_N$.

Next, we define a complex vector bundle over $M$. We take
$U_\alpha\times\C^N$ as the local charts and patch them together
according to the following prescription:
    \begin{enumerate}
    \item For each $(\alpha,\beta)\in\fI^2$ satisfying
    $U_\alpha\cap U_\beta\neq\varnothing$, we define
    $g_{\alpha\beta}:U_\alpha\cap U_\beta\to GL(N,\C)$ by
    \be
    \forall R\in U_\alpha\cap U_\beta,~~~~~~~~
    g_{\alpha\beta}(R):=g(\sigma_R^{\alpha,\beta}).
    \label{LT}
    \ee
    \item For all $R\in U_\alpha\cap U_\beta$ and $\vec z\in\C^N$,
    we identify the point $(R,\vec z)$ in $U_\beta\times\C^N$
    with the point $(R,g_{\alpha\beta}(R)\vec z)$ in
    $U_\alpha\times\C^N$.
    \end{enumerate}
By construction, this yields a fiber bundle $V$ with base manifold
$M$, typical fiber $\C^N$, and structure group $GL(N,\C)$,
\cite{steenrod}. For every $p\in V$, there are $U_\alpha$, $R\in
U_\alpha$, and $\vec z^\alpha\in\C^N$ such that $p$ is represented
by $(R,\vec z^\alpha)$. The projection map $\Pi:V\to M$ acts as
$\Pi(p):=R$, and for each $\alpha\in\fI$ the function
$\Phi_\alpha:\Pi^{-1}(U_\alpha)\to U_\alpha\times\C^N$ defined by
    \[\forall p\in \Pi^{-1}(U_\alpha),~~~\Phi_\alpha(p):=
    (R,\vec z^\alpha),\]
provides a local trivialization of $V$, namely
$(U_\alpha,\Phi_\alpha)$. The functions $g_{\alpha\beta}$ introduced
in (\ref{LT}) are the transition functions associated with these
local trivializations.

Next, for every $\alpha,\beta\in\fI$ fulfilling $U_\alpha\cap
U_\beta\neq\varnothing$, all $R\in U_\alpha\cap U_\beta$ and all
$k\in I_N$, we identify the point $(R,k)$ in $U_\beta\times I_N$
with the point $(R,\sigma_R^{\alpha,\beta}(k))$ in $U_\alpha\times
I_N$. This defines a fiber bundle that we denote by $\cM$. Let
$\bpi:\cM\to M$ be the projection map for $\cM$ and $\cR\in\cM$ be
represented by $(R,k)\in U_\alpha\times I_N$. Then $\bpi(\cR)=R$,
the functions $\phi_\alpha:\bpi^{-1}(U_\alpha)\to U_\alpha\times
I_N$ defined by
    \[\forall p\in \bpi^{-1}(U_\alpha),~~~\phi_\alpha(\cR)=(R,k),\]
give a set of local trivializations $(U_\alpha,\phi_\alpha)$ of
$\cM$, and $\fg_{\alpha\beta}:U_\alpha\cap U_\beta\to \fS_N$ defined
by
    \be
    \forall R\in U_\alpha\cap U_\beta,~~~~~
    \fg_{\alpha\beta}(R):=\sigma_R^{\alpha,\beta},
    \label{LT-2}
    \ee
are the corresponding transition functions.

As a simple example consider the $2\times 2$ Hamiltonians of the
form
    \be
    H=\left(\begin{array}{cc}
    0 & 1\\
    z & 0\end{array}\right),~~~~~~z\in\C-\{0\}.
    \label{H1=}
    \ee
We can conveniently parameterize $H$ by the polar coordinates
$r:=|z|\in\R^+$ and $\theta:={\rm arg}(z)\in\R$. The role of the
manifold $M$ is now played by the punctured plane $\R^2-\{0\}$. The
eigenvalues of $H$ are double-valued functions on $M=\R^2-\{0\}$:
    \be
    E_1=-\sqrt z=-\sqrt r\; e^{\frac{i\theta}{2}},~~~~~
    E_2=\sqrt z=\sqrt r\; e^{\frac{i\theta}{2}}.
    \label{eg-va1=}
    \ee
We can cover $M$ by an open covering consisting of the following
three open subsets.
    \bea
    U_1&:=&\left\{\:r\,e^{i\theta}~|~r\in\R^+,~\theta\in\mbox{$
    (-\frac{\pi}{2},\frac{\pi}{2})$}~\right\},
    \label{oc1}\\
    U_2&:=&\left\{\:r\,e^{i\theta}~|~r\in\R^+,~\theta\in\mbox{$
    (\frac{\pi}{6},\frac{7\pi}{6})$}~\right\},
    \label{oc2}\\
    U_3&:=&\left\{\:r\,e^{i\theta}~|~r\in\R^+,~\theta\in\mbox{$
    (\frac{5\pi}{6},\frac{11\pi}{6})$}~\right\}.
    \label{oc3}
    \eea
The parameter space $\cM$ is the double covering space
\cite{spanier} of $M$ that is constructed as a fiber bundle with
typical fiber $I_2 =\{1,2\}$ and the following transition functions
associated with the open cover $\{U_1,U_2,U_3\}$.
    \bea
    &&\forall R\in U_1\cap U_2
    =\left\{\:r\,e^{i\theta}~|~r\in\R^+,~\theta\in\mbox{$
    (\frac{\pi}{6},\frac{\pi}{2})$}~\right\},
    ~~~~~~~~~~~\fg_{12}(R):={\rm Id},\nn\\
    &&\forall R\in U_2\cap U_3
    =\left\{\:r\,e^{i\theta}~|~r\in\R^+,~\theta\in\mbox{$
    (\frac{5\pi}{6},\frac{7\pi}{6})$}~\right\},
    ~~~~~~~~~\fg_{23}(R):={\rm Id},\nn\\
    &&\forall R\in U_3\cap U_1
    =\left\{\:r\,e^{i\theta}~|~r\in\R^+,~\theta\in\mbox{$
    (-\frac{\pi}{2},-\frac{\pi}{6})$}~\right\},
    ~~~~~~~\fg_{31}(R):=\tau,\nn
    \eea
where ${\rm Id}$ is the identity permutation, ${\rm Id}(k)=k$ for
all $k\in I_2$, and $\tau$ is the transposition that swaps $1$ and
$2$. It is not difficult to identify $\cM$ with the double-sheeted
Riemann surface over which the square root $\sqrt z$ is a
single-valued function of $z$.

In general the situation is more complicated than the simple example
we just considered. As the following example shows, in general there
may exist closed curves in $\cM$ along which some of the eigenvalues
of the Hamiltonian are multiple-valued. Let
    \be
    H[z]:=\left(\begin{array}{cc}
    z & 0\\
    0 & H_{1}[z]\end{array}\right),~~~~H_{1}[z]:=
    \left(\begin{array}{cc}
    0 & 1\\
    z & 0\end{array}\right),~~~~z\in M:=\C-\{0,\pm 1\}.
    \label{H2=}
    \ee
The parameter space $\cM$ for the Hamiltonian $H[z]$ is a three-fold
covering space of $M$. It consists of two connected components that
we denote by $\cM_{1}$ and $\cM_{2,3}$. $\cM_{1}$ is a diffeomorphic
copy of $M$, and $\cM_{2,3}:=\cM'-\bpi^{-1}\{\pm 1\}$ where $\cM'$
is a copy of the parameter space of the Hamiltonian (\ref{H1=}).
Clearly the eigenvalue $E_{1}[z]:=z$ is a single-valued function on
$\cM_1$ and $\cM$, and the eigenvalues $E_{2,3}[z]:=\pm \sqrt{z}$
are single-valued on $\cM_{2,3}$. But the latter are not
single-valued on $\cM$. To see this let $C_1:[0,2\pi]\to M$ be the
curve defined by $z(t)=2e^{it}$, for all $t\in[0,2\pi]$ and
$\cC_1:[0,2\pi]\to\cM$ be the curve contained in $\cM_1$ that
projects onto $C_1$ under $\bpi$. $\cC_1$ is a closed curve in
$\cM$. Clearly $E_1[z(2\pi)]=E_1[z(0)]$, but
$E_{2,3}[z(2\pi)]=-E_{2,3}[z(0)]\neq E_{2,3}[z(0)]$. Therefore,
unlike $\lambda_{1}^\alpha[z(0)]$, the eigenstates
$\lambda_{2,3}^\alpha[z(0)]$ are not adiabatic cyclic states of
period $2\pi$.\footnote{Here $\alpha=1$, if we use the open cover
(\ref{oc1}) -- (\ref{oc3}) of $M$.}

In general, the parameter space $\cM$ consists of up to $N$
connected components;
    $\cM=\cM_1\sqcup\cM_2\sqcup \cdots\sqcup \cM_{K},$
where $K\in I_N$ and $\cM_1$, $\cM_2$, $\cdots$, $\cM_{K}$ are
connected, mutually disjoint, and generally distinct covering spaces
of $M$. For each $j\in I_N$, there is at least one $J\in I_K$ such
that $E_j$ is a single-valued function on $\cM_J$. In other words,
given an initial eigenstate $\lambda_j$, one can choose a
corresponding connected components $\cM_J$ of $\cM$ and use it as
the parameter space for the corresponding geometric phase problem.
This construction has the disadvantage that it makes the parameter
space depend on the spectral label $j$ associated with the initial
state. If we choose $\cM$ as the parameter space for the problem, we
should note that given a closed curve in $\cM$, there may be initial
eigenstates that are not cyclic along this curve. To avoid these
complications we shall next offer an alternative construction for a
parameter space that is more universal in the sense that along every
closed curve in this space all the initial eigenstates undergo
adiabatic cyclic evolutions. These are the very properties of the
parameter space used in Simon's holonomy interpretation of the
adiabatic geometric phase for Hermitian Hamiltonians
\cite{gp-book,simon}.

First, we construct a principle $\fS_N$-bundle over $M$. Let
$\{(U_\alpha,\phi_\alpha)\}_{\alpha\in\fI}$ and $\fg_{\alpha\beta}$
be respectively the family of local trivializations and the
corresponding transition functions that we employed in the
construction of $\cM$. For each $\alpha,\beta\in\fI$ fulfilling
$U_\alpha\cap U_\beta\neq\varnothing$, $R\in U_\alpha\cap U_\beta$,
and $\fg\in\fS_N$, we identify $(R,\fg)\in U_\alpha\times\fS_N$ with
$(R,\fg_{\beta\alpha}(R)\fg)\in U_\beta\times\fS_N$. This defines a
principle $\fS_N$-bundle over $M$ that we denote by  $\tilde\fM$. By
construction, $\cM$ is an associated fiber bundle for $\tilde\fM$,
\cite{Choquet}. $\tilde\fM$ is indeed an $N!$-fold covering space of
$M$ over which the eigenvalues $E_j$ are single-valued functions and
the eigenstates undergo adiabatic cyclic evolutions along of every
closed curve $\tilde\fC:[0,T]\to\tilde\fM$. However, similarly to
$\cM$, $\tilde\fM$ may generally be disconnected. Whenever the
latter is the case, we can find a family of local trivializations of
$\tilde\fM$ whose transition functions belong to a proper subgroup
$\fH$ of the permutation group $\fS_N$, i.e., the structure group
$\fS_N$ is reducible to $\fH$, \cite{Choquet}. This in turn enables
us to reduce $\tilde\fM$ to a principle $\fH$-bundle over $M$.
Choosing the smallest subgroup $\fH$ of $\fS_N$ that allows for the
reduction of $\tilde\fM$, we find an irreducible principle
$\fH$-bundle over $M$ that we label as $\fM$. This is a connected
$|\fH|$-fold\footnote{For a finite set $S$, we use $|S|$ to denote
its order.} covering space of $M$ on which all the eigenvalues of
the Hamiltonian are single-valued.

Computing $\tilde{\fM}$ and $\fM$ for the parameter-dependent
Hamiltonian (\ref{H1=}), we find that they are (diffeomorphic)
copies of the manifold $\cM$ that we obtained above. In contrast,
for the Hamiltonian (\ref{H2=}), $\fM=\cM_{2,3}\varsubsetneq\cM$ and
$\tilde{\fM}=\fM\times I_2$ (up to diffeomorphisms).

If the parameter-dependent Hamiltonian $H[R]$ is Hermitian for all
$R\in M$, the eigenvalues are single-valued functions on $M$ and all
the fiber bundles we have constructed are trivial. In particular, up
to diffeomorphisms, we have $\cM=M\times I_N$, $\tilde{\fM}=M\times
\fS_N$, and $\fM=M$. The latter shows that $\fM$ reduces to the
parameter space for the adiabatic geometric phases associated with
the Hermitian Hamiltonians \cite{gp-book,simon}.

The above properties of $\fM$ make it into the desired parameter
space  for general, possibly non-Hermitian, matrix Hamiltonians.

\section{Construction of the Berry-Simon Line Bundle}

In this section we examine the adiabatic geometric phase
corresponding to the adiabatic cyclic evolution of the eigenstates
of the initial Hamiltonian along an arbitrary closed curve
$\fC:[0,T]\to\fM$ in the parameter space $\fM$.\footnote{Without
loss of generality we can replace $\fM$ with $\tilde{\fM}$ or the
connected component of $\cM$ over which the eigenvalue $E_{n}$ is
single-valued.} Though the eigenstates $\lambda_n$ are single-valued
smooth functions on $\fM$, the eigenvectors $|\psi_n[\cdot]\kt$ can
generally be defined as smooth single-valued functions on certain
open subsets of $\fM$. Let $\{\cU_\fa\}_{\fa\in\fA}$ be an open
cover of $\fM$ such that $\cU_\fa$ are diffeomorphic to $\R^d$ and
$|\psi_n[\cdot]\kt$ be a smooth single-valued function on $\cU_\fa$
for all $\fa\in\fA$. We shall use $\cR:=(\cR_1,\cR_2,\cdots,\cR_d)$
to denote the local coordinate representation of the points of $\fM$
that lie in $\cU_\fa$ and use the superscript $\fa$, $\fb$,~$\cdots$
to denote the local values of the relevant quantities at the points
belonging to $\cU_\fa$, e.g., we denote by $|\psi_n^\fa[\cR(t)]\kt$
the value of $|\psi_n^\fa[\fC(t)]\kt$ for all $t\in[0,T]$ such that
$\fC(t)\in\cU_\fa$.

Suppose for simplicity that $\fC$ lies entirely in $\cU_\fa$ for
some $\fa\in\fA$, in which case we can suppress the coordinate patch
label $\fa$, and consider solving the Schr\"odinger equation
(\ref{scheq}) with the initial condition
    \bea
    |\Psi(t=0)\rangle=k~|\psi_{n}[\cR(0)]\rangle
    \label{2.2},
    \eea
where $k\in\C-\{0\}$ is an arbitrary constant. If we employ
adiabatic approximation, we can express the evolving state vector as
    \be
    |\Psi(t)\kt=k~e^{i[\delta_{n}(t)+\gamma_{n}(t)]}
    |\psi_{n}\left[\mathcal{R}(t)\right]\kt ,
    \label{sol-sch-eq}
    \ee
where
    \be
    \delta_{n}(t):=-\int_{0}^{t}E_{n}(\tau)d\tau,~~~~~~~~
    \gamma_{n}(t):=\int_{0}^{t}i\langle
    \phi_{n}(\tau)|\frac{d}{d\tau}|\psi_{n} (\tau)\rangle\ d\tau,
    \label{2.3}
    \ee
$E_n(t):=E_n[\cR(t)]$, $|\psi_n(t)\kt:=|\psi_n[\cR(t)]\kt$, and
$|\phi_n(t)\kt:=|\phi_n[\cR(t)]\kt$. $\delta_{n}(T)$ and
$\gamma_{n}(T)$ are the dynamical and geometric phase angles
associated with the cyclic evolution of the initial eigenstate
$\lambda_n[\cR(0)]$.

Since $\fM$ is a covering space of $M$, the local properties of the
eigenvectors, eigenvalues, and eigenstates of $H$ as functions on
$\fM$ are identical with their local properties as functions on $M$.
This allows for a straightforward generalization of the Berry-Simon
complex line bundle \cite{gp-book,simon} for non-Hermitian
Hamiltonians. This is the spectral line bundle $L^{n}$ over the
parameter space $\fM$ whose fiber over $\cR\in\fM$ is the eigenstate
$\lambda_n[\cR]\subseteq\cH$. We can construct it as follows.

For each $\fa,\fb\in\fA$ for which
$\cU_\fa\cap\cU_\fb\neq\varnothing$ and for all
$\cR\in\cU_\fa\cap\cU_\fb$, both $|\psi_n^\fa[\cR]\kt$ and
$|\psi_n^\fb[\cR]\kt$ belong to $\lambda_n[\cR]\kt$. Therefore,
there must exist $\cG_{\fa\fb}:\cU_\fa\cap\cU_\fb\to GL(1,\C)$ such
that
    \be
    |\psi_{n}^{\fa}\left[\cR\right]\kt=\cG_{\fa\fb}(\cR)
    |\psi_{n}^{\fb}\left[\cR\right]\kt.
    \label{2.8a}
    \ee
The line bundle $L^{n}$ is obtained by identifying $(\cR,z^\fa)\in
\cU_\fa\times\C$ with $(\cR,\cG_{\fb\fa}(\cR)z^\fa)\in
\cU_\fb\times\C$ for all $\cR\in\cU_\fa\cap\cU_\fb$ and
$z^\fa\in\C$. Let $\fP:L^n\to\fM$ denote the projection map for
$L^n$ and for each $\fa\in\fA$,
$\varphi_\fa:\fP^{-1}(\cU_\fa)\to\cU_\fa\times\C$ be defined by
$\varphi_\fa(\fp):=(\fP(\fp),z^\fa)$ for all $\fp\in
\fP^{-1}(\cU_\fa)$. Then $\{(\cU_\fa,\varphi_\fa)\}_{\fa\in\fA}$ is
a family of local trivializations of $L^n$ with transition functions
$\cG_{\fa\fb}$. This completes the construction of $L^n$ as a
topological line bundle. Next, we endow this line bundle with a
geometric structure, namely Berry's connection one-form whose local
expression is as follows.
    \bea
    \mathcal{A}_{n}^{\fa}(\cR):=i\sum_{j=1}^{d}\br\phi^{\fa}_{n}[\cR]
    |\frac{\partial}{\partial \cR^{j}} | \psi^{\fa}_{n}[\cR]\kt
    d\cR^{j}=i\br \phi^{\fa}_{n}[\cR]|d|\psi^{\fa}_{n}[\cR]\kt.
    \label{2.11}
    \eea
With this choice for the geometry of $L^n$, the geometric phase
factor $e^{i\gamma_n(T)}$ is identified with the holonomy associated
with the curve $\fC$;
    \be
    e^{i\gamma_n(T)}= e^{i\oint_{\fC}\mathcal{A}_{n}^{\fa}}.
    \label{holonomy=}
    \ee
As shown in Appendix~B, this construction applies also to the cases
where $\fC$ does not lie in a single patch $\cU_\fa$.\footnote{In
this construction we can relax the condition that $\cU_\fa$ are
diffeomorphic to $\R^d$. We can work with any open cover
$\{\cU_\fa\}_{\fa\in\fA}$ of $\fM$ such that each $\cU_\fa$ admits
an open cover $\{\fu^\fa_{\beta}\}_{\beta\in\fB}$ with
$\fu^\fa_{\beta}$ diffeomorphic to $\R^d$ for all $\beta\in\fB$.}

We end this section with three remarks.
    \begin{enumerate}

    \item Whenever the Hamiltonian is non-Hermitian, the geometric
    phase factor~(\ref{holonomy=}) is not necessarily unimodular. This
    is consistent with the fact that in general under a non-unitary
    evolution the norm of the evolving state vector changes.

    \item For cases where $H[\cR]$ is Hermitian for all $\cR\in\fM$,
    we can (locally) find an orthonormal set of eigenvectors
    $|\epsilon^\fa_n[\cR]\kt$ of $H[\cR]$ and a set of nowhere vanishing
    functions $\kappa^\fa_n:\cU_\fa\to\C$ such that
    $|\psi_n^\fa[\cR]\kt=\kappa_n^\fa[\cR]|\epsilon^\fa_n[\cR]\kt$,
    $|\phi_n^\fa[\cR]\kt=\kappa_n^\fa[\cR]^{-1*}|\epsilon^\fa_n[\cR]\kt$, and
    the Berry's connection one-form (\ref{2.11}) takes the form
        \[ \cA_{n}^{\fa}(\cR)=i\br
        \epsilon^{\fa}_{n}[\cR]|d|\epsilon^{\fa}_{n}[\cR]\kt+
        id\ln(\kappa_n^\fa[\cR]).\]
    The imaginary part of the first term on the right-hand side of
    this equation vanishes identically:
        \[ \Im\{i\br
        \epsilon^{\fa}_{n}[\cR]|d|\epsilon^{\fa}_{n}[\cR]\kt\}=
        \frac{1}{2i}\left(i\br\epsilon_{n}^{\fa}[\cR]|d\epsilon_{n}^{\fa}[\cR]\kt-
        \{i\br\epsilon_{n}^{\fa}[\cR]|d\epsilon_{n}^{\fa}[\cR]\kt
        \}^*\right)=\frac{1}{2}d\{\br\epsilon_{n}^{\fa}[\cR]
        |\epsilon_{n}^{\fa}[\cR]\kt\}=0.\]
    This shows that for Hermitian Hamiltonians the
    imaginary part of Berry's connection one-form (\ref{2.11}) is an
    exact form (pure gauge). Therefore, it does not contribute to
    the geometric phase factor (\ref{holonomy=}). This is
    consistent with the fact that for Hermitian Hamiltonians the adiabatic
    geometric phase is a genuine (unimodular) phase factor.

    \item Similarly to the case of Hermitian
    Hamiltonians, we can compute the local curvature two-form $F_n$ associated with Berry's
    connection one-form ~(\ref{2.11}). By definition,
        \[ F_n(\cR):= d \cA_n^{\fa}(\cR)=
        i\br d\phi^{\fa}_{n}[\cR]|\wedge|d\psi^{\fa}_{n}[\cR]\kt.\]
    It is easy to check using (\ref{biortho}) that $F^n(\cR)$ is
    invariant under changes of local eigenvectors (gauge transformations).
    Following essentially the same analysis as the one given in
    \cite{gp-book} for Hermitian Hamiltonians, we can put $F^n(\cR)$ in
    the following manifestly gauge-invariant form
    \cite{miniatura,dattoli}.
        \[F_n(\cR)=i\sum_{m\neq n}\frac{\br
        \phi^{\fa}_{n}[\cR]|dH[\cR]|\psi^{\fa}_{m}[\cR]\kt
        \wedge\br \phi^{\fa}_{m}[\cR]|dH[\cR]|\psi^{\fa}_{n}[\cR]\kt}
        {(E_m[\cR]-E_{n}[\cR])^2}.\]
    This expression is also invariant under changes of local trivialization.
    Therefore, it defines $F_n$ as a globally single-valued two-form
    on $\fM$.
    \end{enumerate}

\section{Adiabatic geometric phase for general
$2\times 2$ Hamiltonians}

$2\times 2$ matrix Hamiltonians provide the simplest nontrivial toy
models in the study of geometric phases. As far as the physical
(measurable) quantities are concerned we can confine our attention
to traceless $2\times 2$ Hamiltonians\footnote{One can transform the
set of all non-Hermitian $2 \times 2$ Hamiltonians to set of
traceless $2\times 2$ non-Hermitian Hamiltonians by canonical
transformations that leave all the physical quantities of the
system, including geometric phases, invariant \cite{nova,jmp1999}.}
    \bea
    H=\left(\begin{array}{cc}
    a & b  \\
    c & -a
    \end{array}\right),
    \label{3.1}
    \eea
where $a,b,c\in\C$. It is not difficult to show that the parameter
space $M$ for these models is given by
    \be
    M:=\{~a,b,c\in \C~|~a^2+bc\neq 0~\}\varsubsetneq\C^3=\R^6,
    \ee
It is a real six-dimensional smooth manifold.

$H$ is a single-valued function on $M$ with eigenvalues\footnote{We
use $\pm$ for the spectral label $n$.}
    \bea
    E_{\pm}=\pm f,~~~~f:=\sqrt{a^{2}+bc}.
    \label{3.2}
    \eea
Because of the appearance of the square root (with a generally
complex argument) in the expression for $f$, $E_\pm$ are, in
general, only locally single-valued functions on $M$. The parameter
space $\fM$, which happens to coincide with $\cM$ and $\tilde\fM$,
is a double covering space of $M$ over which $f$ is a single-valued
function. This implies that each point of $\fM$ is uniquely
determined by the values of $a,b,c$ and $f$. Note that $a,b,c$ does
determine $f$ up to a sign. Specifying the value of $f$ corresponds
to fixing the sign ambiguity. In the following we shall use $\cR$ to
denote the local coordinates of the points of $\fM$; $\cR$ is
determined by the valued of $a,b,c$ and $f$.

We can view the eigenvectors $|\psi_\pm[\cdot]\kt$ of $H$ as global
sections of the line bundles $L^{\pm}$. It turns out that these fail
to be continuous global sections, and that we can only construct
smooth local sections corresponding to $|\psi_\pm[\cdot]\kt$. This
requires constructing an open cover of $\fM$. To achieve this, we
introduce
    \bea
    \cP^{\pm}:=\left\{ \cR \in\fM~|~~ bc=0,~~f=\pm a\neq 0\right\},
    \label{3.4}
    \eea
which are a pair of disjoint closed subsets\footnote{Note that on
$\fM$, $f\neq 0$. This is the reason why $\cP^\pm$ are closed
subsets of $\fM$.} of $\fM$, and consider their complement in $\fM$,
namely
    \bea
    \fM^{1}:=\fM-\cP^{-},~~~~\fM^{2}:=\fM-\cP^{+}.
    \label{3.5}
    \eea
Clearly these are open subsets of $\fM$ satisfying
$\fM^{1}\cup\fM^{2}=\fM$. Hence $\{\fM^1,\fM^2\}$ is an open cover
of $\fM$.

Now, we are in a position to define a set of eigenvectors of the
Hamiltonian (\ref{3.1}) that yield smooth local sections of the
bundle $L^\pm$. These are the functions
$|\psi_\pm^\nu[\cdot]\kt:\fM^\nu\to L^\pm$, with $\nu\in I_2$, that
are defined by
    \be
    |\psi_{+}^{1}[\cR]\rangle :=\left(\begin{array}{c}
    f+a\\
    c
    \end{array}\right),~~~~
    |\psi_{-}^{1}[\cR]\rangle :=\left(\begin{array}{c}
    -b\\
    f+a
    \end{array}\right),~~~~
    |\psi_{\pm}^{2}[\cR]\kt:= |\psi_{\mp}^{1}[\cR]\kt\Big|_{f\to-f}.
    \label{3.6}
    \ee
We can also construct the corresponding left eigenvectors
$|\phi^\nu_\pm[\cR]\kt$ of the Hamiltonian that together with
$|\psi^\nu_\pm[\cR]\kt$ form a biorthonormal system for the Hilbert
space:{\small
    \be
    |\phi_{+}^{1}[\cR]\rangle =\frac{1}{2f^{*}(f^{*}+a^{*})}
    \left(\!\begin{array}{c}
    \!f^{*}+a^{*}\!  \\
    b^{*}
    \end{array}\!\right),~~~
    |\phi_{-}^{1}[\cR]\rangle=\frac{1}{2f^{*}(f^{*}+a^{*})}
    \left(\!\begin{array}{c}
    -c^{*}  \\
    \!f^{*}+a^{*}\!
    \end{array}\!\right),~~~
    |\phi_{\pm}^{2}[\cR]\kt:= |\phi_{\mp}^{1}[\cR]\kt\Big|_{\!f\to-f}.
    \label{3.6b}
    \ee}

Next, we use (\ref{3.6}) and (\ref{3.6b}) to give local expressions
for Berry's connection one-form on $L^\pm$. Using the superscript
$\nu$ to emphasize that the domain of definition of the
corresponding connection one-form is $\fM^\nu$, we have
    \bea
    \mathcal{A}^{1}_{+}(\cR) &=& \frac{i}{2f}
    \left( \frac{bdc}{f+a}+df+da \right),~~~~
    \mathcal{A}^{1}_{-}(\cR) = \frac{i}{2f}\left(
    \frac{cdb}{f+a}+df+da \right),
    \label{3.7}\\
    \mathcal{A}^{2}_{\pm}(\cR)&=&
    \mathcal{A}^{1}_{\mp}(\cR)\Big|_{f\to-f}.
    \label{3.7b}
    \eea
For a closed curve $\fC:[0,T]\to\fM$ that lies entirely in $\fM^1$
or $\fM^2$ we can use (\ref{holonomy=}) and either of (\ref{3.7}) or
(\ref{3.7b}) to calculate the corresponding adiabatic geometric
phase. If $\fC$ is not contained in $\fM^1$ or $\fM^2$, we must
compute the contributions to the geometric phase from segments of
$\fC$ that lie in $\fM^1$ and $\fM^2$ and use the transition
functions $\cG^{\pm}_{2,1}:\fM^1\cap\fM^2\to GL(1,\C)$ to patch up
these contributions and obtain the total geometric phase. We give
the details of this well-known construction in Appendix~B. Here we
only include the expression for the transition functions that we
obtained using (\ref{2.8a}) and (\ref{3.6}):
    \bea
    \cG^{+}_{2,1}(\cR)= \frac{-b}{f+a},~~~~\cG^{-}_{2,1}(\cR)=
    \frac{c}{f+a},~~~~ \forall \cR \in \fM^{1}\cap\fM^{2}.
    \label{3.10}
    \eea

In the remainder of this section we examine a number of specific
examples. These will in particular be helpful in our discussion of
the geometrical versus topological nature of EP-related geometric
phases.

\subsection{Two symmetric examples}

In this subsection we determine the parameter space $\fM$ for two
symmetric matrix Hamiltonians whose EPs are branch points of the
function $f$ over the parameter space $M$. We also give the
adiabatic geometric phase for curves enclosing the EPs.

\begin{itemize}
\item[a.] Consider the symmetric Hamiltonian
    $$  H[z] :=\left(\begin{array}{cc}
    1+z & i(1-z)  \\
    i(1-z) & -( 1+z )
   \end{array}\right),~~~z\in M:= \C-\{0\},$$
for which $f=2\sqrt{z}$. The parameter space $\fM$ is a double
covering space of $M$ over which $f$ is single-valued. Clearly, it
is the two-sheeted Riemann surface over the punctured complex plane
which we described in Section~2. The point $z=0$ is the only EP for
this Hamiltonian.

\item[b.] Let
    $$  H[z] :=\left(\begin{array}{cc}
    1+z & 1-z  \\
    1-z & -( 1+z )
   \end{array}\right),~~~z\in M:= \C-\{- i,+i\}.$$
Then $f=2\sqrt{(1+z^{2})}=2\sqrt{(z+i)(z-i)}$ and $\fM$ is the
Riemann surface with two branched points \cite{churchill}.
\end{itemize}
For both of these examples, the Hamiltonian is symmetric, the
Berry's connection one-form is flat, and consequently the geometric
phase factor associated with any closed curve encircling an EP has
the topological value of $\pm 1$, \cite{mail1}.

\subsection{Two non-Symmetric examples}

Here we consider a pair of non-symmetric Hamiltonians depending also
on a single complex parameter and examine the adiabatic geometrical
phase problem for closed curves enclosing EPs.

\begin{itemize}
\item[a.] For the Hamiltonian
    \be
    H[z] :=\left(\begin{array}{cc}  z & 1  \\
    0 & -z
    \end{array}\right),~~~~z\in M:=\C-\{0\},
    \label{H2a}
    \ee
we have $a=f=z$, $b=1$, and $c=0$. Therefore, $z=0$ is an EP,
$\tilde{\fM}=\cM=\{-1,+1\}\times M$, $\fM=M$, and in view of
Eqs.~(\ref{3.6}) -- (\ref{3.7b}),
    $$\mathcal{A}_{+}(\cR)=0,~~~~\mathcal{A}_{-}(\cR)=\frac{i\,dz}{z}.$$
This together with (\ref{holonomy=}) yield the value of $0$ for the
adiabatic geometric phase angle corresponding to every closed curve
in $\fM$ including those encircling the EP.

\item[b.] Consider the following family of Hamiltonians
    $$ H[z] =\left(\begin{array}{cc}
    \alpha z & 1  \\
    (\beta^{2}-\alpha^{2})z^{2}& -\alpha z
    \end{array}\right), $$
where we treat $\alpha$ and $\beta$ as fixed but arbitrary nonzero
complex constants and $z$ as the physical parameter that is made
time-dependent. Then $z=0$ is the only EP, $M:=\C-\{0\}$, $f=\beta
z$, $\fM=M$, and in light of (\ref{3.6}) -- (\ref{3.7b}),
    $$\mathcal{A}_+(\cR)=
    i\frac{dz}{z}\left(\frac{3\beta-\alpha}{2\beta} \right),
    ~~~~~\mathcal{A}_-(\cR)=i\frac{dz}{z}
    \left(\frac{\alpha+\beta}{2\beta} \right).$$
For this system the adiabatic geometric phase angles
$\gamma_{\pm}(T)$ associated with the curve $\fC(t):=e^{2\pi it/T}$
are given  (up to integer multiples of $2\pi$) by
    $$\gamma_{+}(T)=-\pi\left(1-\frac{\alpha}{\beta}
    \right),~~~~
    \gamma_{-}(T)=-\pi\left(1+\frac{\alpha}{\beta}\right).$$
Note that $\fC$ does enclose the EP, but depending on the choice of
$\alpha$ and $\beta$, $\gamma_{\pm}(T)$ can take arbitrary values.
This seems to contradict the results of \cite{mail1} where the
adiabatic geometric phase angles for general $2\times 2$
Hamiltonians are claimed to take the values $\pm\pi$.
\end{itemize}

\subsection{A three-parameter family of Hamiltonians}

In Ref.~\cite{mondragon}, the authors consider traceless $2\times 2$
Hamiltonians with constant non-Hermitian part, i.e.,
$H[R]=(\vec{R}-\frac{i}{2}\vec{\Gamma}).\vec{\sigma}$, where $\vec
R:=R=(R_1,R_2,R_3)\in\R^3$ are the physical parameters of the
system, $\vec{\Gamma}=(\Gamma_1,\Gamma_2,\Gamma_3)\in \R^3-\{\vec
0\}$ is a constant nonzero vector, $\vec{\sigma}:=(\sigma_1,
\sigma_2,\sigma_3)$, and $\sigma_1, \sigma_2,\sigma_3$ are Pauli
matrices. Because, we can transform the anti-Hermitian part of
$H[R]$, namely $\frac{i}{2}\vec{\Gamma}.\vec{\sigma}$, to a diagonal
matrix by performing a constant similarity transformation, we can
confine our attention to the case $\vec{\Gamma}=(0,0,\Gamma_3)$ with
no lose of generality. This leads to the following explicit form for
$H[R]$.
    \be
    H[R] =\left(\begin{array}{cc}
    R_3-\frac{i\,\Gamma}{2} & R_1-iR_2  \\
    R_1+iR_2& -( R_3-\frac{i\,\Gamma}{2} )
    \end{array}\right),
    \label{e3}
    \ee
where we dropped the label $3$ in $\Gamma_3$. It is easy to see that
$f=\sqrt{|\vec{R}|^2-iR_3\Gamma-\frac{\Gamma^2}{4}}$. Therefore,
there is an infinity of EPs forming the circle:
    \[ S^1:=\left\{(R_1,R_2,R_3)\in\R^3~|~R_1^2+R_2^2=
    \mbox{$\frac{\Gamma^2}{4}$},~R_3=0~\right\}.\]
This implies that $M:=\R^3-S^1$ and $\fM$ is a double covering space
of $M$ over which $f$ is single-valued.

The authors of Refs.~\cite{mondragon} and \cite{nesterov} calculate
the geometric phase for closed curves in $M$ (not $\fM$) that
encircle or approach $S^1$. If one changes the parameters along the
non-contractible curves $C$ in $M$ (by traversing it once), the
eigenstates of the Hamiltonian do not return to their initial value.
Therefore, they do not undergo (adiabatic) cyclic evolutions. In the
following, we examine the structure of the parameter space $\fM$ for
this model.

We can simplify the situation by considering the subfamily of
(\ref{e3}) obtained by setting $R_{2}=0$. Then
$f=\sqrt{R_1^2+R_3^2-iR_3\Gamma-\frac{\Gamma^2}{4}}$, the
Hamiltonian $H[R]$ (with $R:=(R_1,R_3)$) has a pair of EPs located
at $R_\pm:=(R_1=\pm \frac{\Gamma}{2},R_3=0)$, and
$M=\R^2-\{R_-,R_+\}$.

Now, consider the closed curves $C_{\pm}:[0,T]\to M$ defined by
    $$C_{\pm}(t)=(R_1(t),R_3(t))=\left(
    \pm(\frac{\Gamma}{2}+\epsilon \cos \omega t),\epsilon\sin\omega
    t\right),~~~~ \forall t\in[0,T],$$
where $\omega:=2\pi/T$ and $\epsilon$ is a positive real number less
than $\Gamma/2$. It is not difficult to see that $C_\pm$ encloses
the EP located at $R_{\pm}$.

Along $C_\pm$ the value of $f$ changes according to
    $$f(C_\pm (t))=\sqrt{\epsilon(\epsilon+\Gamma e^{-i\omega t})}.$$
Hence $f(C_\pm (T))=-f(C_\pm(0))$. The same holds for every closed
curve in $M$ that is homotopic to $C_\pm$. Furthermore, because $f$
behaves similarly to the complex-valued function
$h(z):=\sqrt{(z-\frac{\Gamma}{2})(z+\frac{\Gamma}{2})}$, the
parameter space $\fM$, over which $f$ is single-valued, is
diffeomorphic to the Riemman surface corresponding to $h(z)$. See
for example \cite{churchill}. To the closed curves $C_\pm$ in $M$
one may associate the closed curves $\fC_\pm$ in $\fM$ that project
under the bundle projection map $\bpi:\fM\to M$ onto $\fC_\pm$, and
as one goes around $\fC_\pm$ once, one traverses $C_\pm$ twice.
Under this change of parameters, the eigenstates of the initial
Hamiltonian perform a cyclic adiabatic evolution, and according to
the analysis of \cite{mail1} the corresponding geometric phases are
topological in nature. As we explain in the next subsection, this
behavior stems from the choice of traversing $C_\pm$ twice.

\subsection{Are EP-related geometric phases topological?}

In Ref.~\cite{mail1}, the authors calculate the adiabatic geometric
phase around the EPs for the general $2\times 2$ matrix Hamiltonians
and find it to be topological, i.e., the geometric phase angle
associated with each cycle equals $\pm\pi$.\footnote{This is in
obvious conflict with the results of Subsection~4.2.} In the
following, we reexamine the argument leading to this conclusion, and
show that it is based on certain implicit assumptions that do not
hold generally.

We begin our assessment of the above-mentioned result of
\cite{mail1} for nonsymmetric $2\times 2$ Hamiltonians by noting
that the authors of \cite{mail1} work with the parameter space $M$
instead of $\fM$. Consequently, in order to make eigenstates undergo
adiabatic cyclic evolutions, they consider encircling EPs along
closed curves $C$ in $M$ twice. They use the notation $2C$ to
emphasize this point. This applies whenever the EP is a branch
point. There are two shortcomings with this approach. Firstly, there
are cases where the enclosed EP is not a branch point. A simple
example is the EP associated with the Hamiltonian (\ref{H2a}).
Secondly, and much more importantly, to obtain an adiabatic cyclic
evolution of the eigenstates of the initial Hamiltonian one does not
need to traverse the same curve $C$ twice. One can alternatively go
around $C$ once and then follow an arbitrary closed curve $C'$ that
does not enclose any EPs other than those enclosed by $C$. We denote
the resulting combined curve by $C+C'$ which yield $2C$ if we choose
$C'=C$ (with the same orientation). More generally, we can choose
(in place of $C+C'$) an arbitrary smooth curve $\tilde C$ in $M$
that is homotopic to $2C$. The domain of the validity of the result
of \cite{mail1} pertaining the topological nature of the geometric
phases is restricted to branch point EPs and the special choice
$\tilde C=2C$. In general, for an EP which is a branch point there
are an infinity of other choices for $\tilde C$ and the resulting
geometric phases are sensitive to its shape. Furthermore, whenever
the EP is not a branch point, the cyclic evolution is generated by
traversing a closed curve in $M$ once. In this case, one obtains the
geometric phase of Garrison and Wright \cite{gari}, which being a
direct generalization of Berry's phase \cite{berry}, is generally
geometrical and not topological.

\section{Conclusion}

If an eigenstate of the Hamiltonian is a single-valued function on a
closed curve $C$ in the space $M$ of the parameters of the
Hamiltonian, it undergoes an adiabatic cyclic evolution along $C$.
In this case, one may pursue the approach of Garrison and Wright and
define the non-Hermitian generalization of Berry's adiabatic
geometric phase for this cyclic evolution. This is however not
generally the case. If $C$ encircles an exceptional point, the
eigenstate in question might not be single-valued along $C$. A
careful description of this situation calls for a closer look at the
appropriate parameter space for the problem of the adiabatic
geometric phase for non-Hermitian Hamiltonians. In this article, we
have constructed an appropriate parameter space $\fM$ for this
problem that is a certain covering space of $M$. We have then
introduced the non-Hermitian generalization of the Berry-Simon
complex line bundle over $\fM$ whose holonomies coincide with the
adiabatic geometric phases.

For $2\times2$ matrix Hamiltonians, we have offered explicit and
general formulas for the adiabatic geometric phase corresponding to
arbitrary closed curves in $\fM$ and showed that contrary to some
recent claims these phases are not generally topological in nature.
In light of the developments reported in
\cite{his-prl,stehmann-jpa-2004}, it should be possible to confirm
this prediction experimentally.

\section*{Acknowledgment}

HMD would like to express his gratitude to Mostafa Esfahanizadeh,
Mohammad Reza Razvan and Mohammad Khorrami for fruitful discussions
and helpful comments.

\begin{appendix}
\section*{Appendices }

\section{Algebraic Topological Properties of $\tilde\fM$ }

In this appendix we use some well-known algebraic topological tools
to explore certain properties of the parameter space $\tilde{\fM}$
which we constructed as a principal $\fS_N$-bundle over $M$ with
projection map $\bpi:\tilde\fM\to M$.

We begin by recalling that each closed curves $C$ in $M$ that passes
through a point $R\in M$ determines an element $[C]$ of the
fundamental group $\pi_{1}(M,R)$ of $M$. Because $M$ is a connected
manifold, we can omit the base point $R$ and use $\pi_{1}(M)$ for
$\pi_{1}(M,R)$, \cite{nakahara}. We can define a right action of $C$
on the fiber $\bpi^{-1}(R)$ over $R$ which maps every point
$\cR_i\in\bpi^{-1}(R)$ to the end point $\cR_{i}.C\in \bpi^{-1}(R)$
of the unique lift of $C$ to $\tilde\fM$ that has $\cR_i$ as its
initial point. Because $\tilde\fM$ is a covering space of $M$,
homotopic curves in $M$ yield the same right action on
$\bpi^{-1}(R)$, \cite{massey,lee}. Therefore, the above construction
defines a right action of the fundamental group $\pi_1(M)$ on the
fibers of $\tilde\fM$.

Next, consider the homomorphism
$\bpi_{*}:\pi_{1}(\tilde{\fM},\cR_{i})\rightarrow \pi_{1}(M)$
induced by $\bpi$ that maps the homotopy class $[\cC_i]$ of the
closed curves $\cC_i$ passing through $\cR_i$ in $\tilde\fM$ onto
the homotopy class $[\bpi\circ\cC_i]$ of the projection of $\cC_i$
under $\bpi$.\footnote{Note that given a connected covering space
$\cN$ of a manifold $N$, an arbitrary point $R\in N$, and any two
points $\cR_{i}$ and $\cR_{j}$ of the fiber over $R$,
$\bpi_{*}(\pi_{1}({\cN},\cR_{i})))$ and
$\bpi_{*}(\pi_{1}({\cN},\cR_{j}))$ are conjugate subgroups of
$\pi_{1}(N)$ \cite{massey,lee}.} One can show that for every
$\cR_{i}\in \bpi^{-1}(R)$, and every curve $C$ in $M$ whose homotopy
class $[C]$ belongs to $\bpi_{*}(\pi_{1}(\tilde{\fM},\cR_{i}))$,
$\cR_i\cdot C=\cR_i$. Therefore,
$\bpi_{*}(\pi_{1}(\tilde{\fM},\cR_{i}))$ has a trivial action on
$\bpi^{-1}(R)$, i.e., with respect to its action on $\bpi^{-1}(R)$,
$\bpi_{*}(\pi_{1}(\tilde{\fM},\cR_{i}))$ is an isotropy subgroup of
$\pi_{1}(M)$. This in turn implies that
$\bpi_{*}(\pi_{1}(\tilde{\fM},\cR_{j}))$ is a normal subgroup of
$\pi_{1}(M)$; for every $[C]\in
\bpi_{*}(\pi_{1}(\tilde{\fM},\cR_{j}))$ and every $[C_{0}]\in
\pi_{1}(M)$ we have $\cR_{i}.(C_{0}.C.C_{0}^{-1})=\cR_{i}.C\in
\tilde{\fM}$. We can use this observation to infer that
$\bpi_{*}(\pi_{1}(\tilde{\fM},\cR_{i}))$ is independent of the
choice of $\cR_i$. Therefore, we shall use the abbreviated notation
$\bpi_{*}(\pi_{1}(\tilde{\fM}))$ for
$\bpi_{*}(\pi_{1}(\tilde{\fM},\cR_{i}))$.

For the parameter space $\cM$ which is also a covering space of $M$,
it may happen that $\bpi_{*}(\pi_{1}({\cM},\cR_{i}))$ depends on
$\cR_{i}$ and $\bpi_{*}(\pi_{1}({\cM},\cR_{i}))$ fails to be an
isotropy subgroup of $\pi_{1}(M)$. In this case there exists closed
curves in $\cM$ that do not yield a cyclic evolution for some of the
eigenstates of the initial Hamiltonian.

It is possible for $\tilde{\fM}$ to be a disconnected covering space
of $M$. In this case, the right action of $\pi_{1}(M)$ on
$\bpi^{-1}(R)$ is not transitive. This means that we can choose the
local trivializations of $\tilde{\fM}$ in such a way that the
transition functions take their values in an irreducible proper
subgroup $\mathfrak{H}$ of $\mathfrak{S}_{N}$. We will next denote
by $\fM$ one of the connected components of $\tilde{\fM}$, and argue
that $\fM$ and every other connected component of $\tilde{\fM}$ is
diffeomorphic as manifolds with the principal $\fH$-bundle obtained
by reducing $\tilde\fM$.

As shown in \cite{massey}, $\bpi^{-1}(R)\cap\fM$ is equivalent
(bijective) to the quotient set $\pi_{1}(M)/\bpi_{*}(\pi_{1}(\fM))$,
and the order of the fibers of $\fM$ is equal to the order of
$\pi_{1}(M)/\bpi_{*}(\pi_{1}(\fM))$. Note that
$\bpi_{*}(\pi_{1}(\fM))$ is a normal subgroup of $\pi_{1}(M)$, and
$\pi_{1}(M)/\bpi_{*}(\pi_{1}(\fM))$ is a group isomorphic to $\fH$.
Therefore, the order of fibers of $\fM$ is equal to the order
$|\mathfrak{H}|$ of $\fH$. Now, consider arbitrary points
$\cR_j,\cR_k\in \bpi^{-1}(R)$ such that $\cR_j\in\fM$ and
$\cR_k\notin\fM$. Since $\cR_k$ is equivalent to an ordered set of
eigenvalues of $H[R]$, the action of the group $\pi_{1}(M)$ on
$\cR_{k}$, is equivalent to its action on $\cR_{j}\in \bpi ^{-1}(R)
\cap \fM$. This suggests the existence of an $|\mathfrak{H}|$-fold
covering space $\fM'$ of $M$ that contains $\cR_{k}$ and is
isomorphic (as bundles) to $\fM$. As a result the parameter space
$\tilde{\fM}$ consists of $N!/|\mathfrak{H}|$ disjoint connected
$|\mathfrak{H}|$-fold covering spaces of $M$ which are diffeomorphic
to $\fM$.

\section{Adiabatic Geometric Phase for a General Closed Curve in $\fM$}

First, consider a closed $\fC$ lying in an open patch $\cU_{\fa}$ of
$\fM$, so that we can suppress the patch label $\fa$ in our
calculations. Then, given an adiabatic solution $|\Psi(t)\kt$ of the
Schr\"{o}dinger equation (\ref{scheq}) fulfilling the initial
condition (\ref{2.2}), the relation
    \bea
    |\tilde{\Psi}[\cR(t)]\kt:=e^{-i\delta_{n}(t)}|\Psi(t)\kt
    :=k\,e^{i\gamma_{n}(\fC(t))}|\psi_{n}\left[\mathcal{R}(t)\right]\kt,
    \label{2.9}
    \eea
defines a single-valued function $|\tilde{\Psi}[\cdot]\kt:\fC\to
L^n$, which we identify with the horizontal lift of the curve $\fC$,
\cite{gp-book,Choquet,isham,nakahara}. As $t$ varies in the range
$[0,T]$, $|\tilde{\Psi}[\cR(t)]\kt$ traces the parallel transport of
$|\Psi(\cR(0))\kt$ along $\fC$. The phase factor $\exp
(i\gamma_{n}(\fC(t)))$ appearing in (\ref{2.9}) is not
gauge-invariant unless $\cR(t)=\cR(0)$. By construction, the latter
holds for $t=T$. The phase factor $\exp (i\gamma_{n}(\fC(T)))$ is
the holonomy of the closed curve $\fC$.

Next, we wish to generalize the expression for $\exp
(i\gamma_{n}(\fC(T)))$ to the cases that $\fC$ does not lie in a
single open patch. This requires the examination of the behavior of
the Berry's connection one-form (\ref{2.11}) under a change of local
trivializations. In view of (\ref{2.8a}) and (\ref{2.11}),
    \bea
    \mathcal{A}_{n}^{\fb}(\cR)=\mathcal{A}_{n}^{\fa}(\cR)
    +i\frac{d \cG_{n}^{\fb,\fa}(\cR)}{\cG_{n}^{\fb,\fa}(\cR)}.
    \label{2.12}
    \eea
Suppose that $\fC$ is a closed curve lying in the union of the open
patches $\cU_0,\cU_1,\cU_2,\cdots,\cU_{r-1}$ where $r\in\Z^+$. See
footnote 9 and Figure 1.a.
    \begin{figure}
    \vspace{0.0cm} \hspace{0.00cm}
    \centerline{\includegraphics[scale=.75,clip]{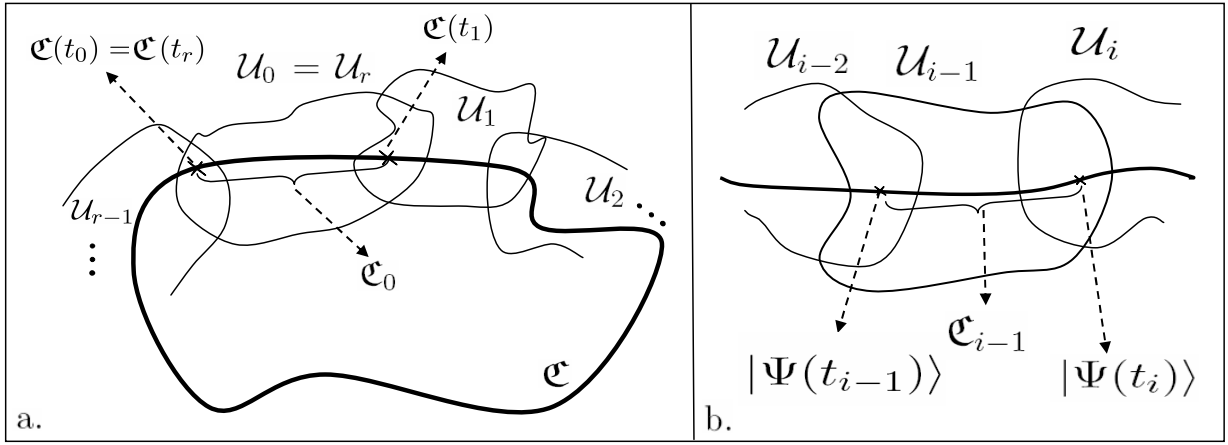}}
    \centerline{\parbox{16cm}{\caption{a.~An arbitrary closed curve $\fC$
    (the bold curve) visiting different open patches.
    b.~The $i-1$'th segment of the curve $\fC$, namely
    $\fC_{i-1}$, with the initial and final points $\fC_{i-1}(t_{i-1})$
    and $\fC_{i-1}(t_{i})$.}}}
    \label{fig1}
    \vspace{0.0cm}
    \end{figure}
Let $\cU_r:=\cU_0$; $t_0,t_1,t_2,\cdots,t_{r-1},t_r\in[0,T]$ be such
that $0=t_0< t_1<t_2<\cdots<t_{r-1}<t_r=T$; for all
$i\in\{1,2,\cdots,r\}$, $\fC(t_i)\in\cU_{i-1}\cap\cU_{i}$,
$\fC(t_0):=\fC(t_r)\in\cU_{r-1}\cap\cU_{0}$ and for all
$t\in[t_{i-1},t_{i}]$ the point $\fC(t)$ belongs to $\cU_{i-1}$. The
latter determine the segments of $\fC$ belonging to $\cU_{i-1}$ that
we denote by $\fC_{i-1}$.\footnote{It can occur that the curve $\fC$
visits a given patch more than once. We account for this situation
by allowing the possibility of using different labels for the same
patch, i.e., $\mathcal{U}_{{i}}$ are not necessarily distinct
patches.} Next, let
$|\psi_{n}^{{i-1}}[\cdot]\kt:\mathcal{U}_{{i-1}}\to L^n$ be the
smooth local section defined on $\mathcal{U}_{{i-1}}$.

For $t\in[t_{i-1},t_{i}]$, the state vector $|\Psi(t)\kt$ that
evolves adiabatically according to the Schr\"odinger equation
(\ref{scheq}) with the initial condition $|\Psi(t_{i-1})\kt=k_{i-1}
|\psi_{n}^{{i-1}}[\cR(t_{i-1})]\kt$, satisfies
    \bea
    |\Psi(t_{i})\kt=k_{i}
    |\psi_{n}^{{i}}[\cR(t_{i})]\kt,
    \label{2.1.3}
    \eea
(See Figure 1.b.). Here the coefficient $k_{i}$ is given by
    \bea
    k_{i}:=\cK_{i}k_{i-1},~~\cK_{i}:=e^{i\left(
    \delta^{i-1}_{n}+\gamma_{n}(\fC_{i-1})\right)}
    \cG_{n}^{{i-1},{i}}(\cR(t_{i})),~~~~k_{0}:=k,
    \label{2.1.4}\\
    \delta^{i-1}_{n}:=-\int_{t_{i-1}}^{t_{i}}E_{n}(t)dt,~~~~
    \gamma_{n}(\fC_{i-1}):=\int_{\fC_{i-1}}
    \mathcal{A}_{n}^{{i-1}},\hspace{1.6 cm}
    \label{2.1.5}
    \eea
and $\cG_{n}^{i-1,i}$ and $\mathcal{A}_{n}^{i-1}$ are respectively
given by (\ref{2.8a}) and (\ref{2.11}). We can use (\ref{2.1.3})
successively by setting $i=1,2,\cdots,r$ to obtain the adiabatic
solution of (\ref{scheq}) with initial condition (\ref{2.2}) as
  \bea
   |\Psi(T)\kt=k_{r}|\psi_{n}^{{r}}[\cR(t_{r})]\kt=
   k_{r}|\psi_{n}^{{0}}[\cR(t_{0})]\kt.
   \label{2.1.6}
   \eea
Furthermore, using (\ref{2.1.4}) and (\ref{2.1.5}) recursively we
find
   \bea
   k_{r}=\cK_{r}k_{r-1}=k_{0}\prod^{r}_{i=1}\cK_{i}
   = k e^{-i\int_{0}^{T} E_{n}(t)dt}\prod_{i=1}^{r}
   \left\{e^{i\left( \gamma_{n}(\fC_{i-1})\right)}
   \cG_{n}^{{i-1},{i}}(\cR(t_{i})) \right\}.
   \label{2.1.7}
   \eea
Note that in this construction the label $i$ is cyclic in the sense
that the label $r$ is to be identified with $0$. For example,
$\cG_{n}^{{r-1},{r}}=\cG_{n}^{{r-1},{0}}$, because
$\mathcal{U}_{{r}}=\mathcal{U}_{{0}}$. The geometric phase
associated with the closed curve $\fC$ has the form
    \bea
    e^{i\gamma_{n}(\fC)}= \prod_{i=1}^{r}
    \left\{e^{i\left( \gamma_{n}(\fC_{i-1})\right)}
    \cG_{n}^{{i-1},{i}}(\cR(t_{i})) \right\}.
    \label{2.1.8}
    \eea

\end{appendix}

\ed